\documentclass[reprint,preprintnumbers,amsmath,amssymb,aps]{revtex4-1}

\usepackage{slashed}
\usepackage{graphicx}
\usepackage{dcolumn}
\usepackage{bm}
\usepackage[dvipsnames]{xcolor}
\usepackage{hyperref}

\def\beq{\begin{equation}\begin{aligned}}
\def\eeq{\end{aligned}\end{equation}}

\newcommand{\be}{\begin{equation}} 
\newcommand{\ee}{\end{equation}}

\begin{document}

\title{\mbox{Comment on ``keV Neutrino Dark Matter in a Fast Expanding Universe'' by Biswas et al.}}
\author{Nicolas~Fernandez}
\affiliation{Department of Physics, 1156 High St., University of California Santa Cruz, Santa Cruz, CA 95064, USA}
\affiliation{Santa Cruz Institute for Particle Physics, 1156 High St., Santa Cruz, CA 95064, USA}

\author{Stefano~Profumo}
\affiliation{Department of Physics, 1156 High St., University of California Santa Cruz, Santa Cruz, CA 95064, USA}
\affiliation{Santa Cruz Institute for Particle Physics, 1156 High St., Santa Cruz, CA 95064, USA}


\begin{abstract}
Biswas et al. \cite{Biswas:2018iny} found that the thermal relic density of a dark matter particle freezing out while the universe's energy density is dominated by a non-standard extra component $\phi$, whose energy density redshifts faster than radiation, can be greatly suppressed. Here we show that this result, which contradicts extensive previous literature, is incorrect: the mistake lies with the assumption that the (decoupled) extra component $\phi$ contributes to the entropic degrees of freedom relevant for dark matter freeze out. If this were the case, a completely different approach would be needed to calculate the dark matter relic abundance, with dramatically different results. 
\end{abstract}

\maketitle

A recent manuscript by Biswas et al. \cite{Biswas:2018iny} studies the thermal relic density of sterile neutrinos in a ``fast-expanding'' universe that includes an additional energy density component from a non-interacting additional component $\phi$ whose energy density red-shifts with the scale factor $a$ as $$\rho_\phi\propto a^{-(4+n)},\ \ n>0.$$ Ref.~\cite{Biswas:2018iny} claims to derive the surprising result that the thermal relic density of sterile neutrinos (or of any other thermal relic) freezing out while the universe's energy density is dominated by $\phi$ is {\em suppressed} compared to the standard case of freeze-out in radiation domination, because of an additional contribution to the entropic degrees of freedom from $\phi$. This would contradict the extensive previous literature on this topic, see e.g. Refs.~\cite{Salati:2002md, Profumo:2003hq, Redmond:2017tja, DEramo:2017gpl}, which indicates that the relic density of particles freezing out when the universe is $\phi$-dominated is always {\em larger} than in the standard case. Here, we show that the result by Biswas et al. \cite{Biswas:2018iny} is incorrect.

The expressions the authors of Ref.~\cite{Biswas:2018iny} use for the number of effective entropy degrees of freedom, Eq.~(30) and (31) in Ref.~\cite{Biswas:2018iny}, are incorrect for the case under consideration. In order to define the entropy of $\phi$ as  $s = (1+w)\rho/T$ (the unnumbered equation between Eq.~(27) and (28) in Ref.~\cite{Biswas:2018iny}), with $T$ the temperature of the standard relativistic thermal bath, $\phi$ must be in local thermal equilibrium with the sector at temperature $T$ (here, ``radiation'', i.e. the standard relativistic thermal bath) with which it shares the entropy: there must be some fast-enough process that thermalizes $\phi$ with the other particles it shares the entropy with. Effectively, by using the equation $s = (1+w)\rho/T$, Ref.~\cite{Biswas:2018iny} assumes that the radiation bath and  $\phi$ are in local thermal equilibrium.

However, if this were the case, and, therefore, if interactions between Standard Model relativistic particles and $\phi$ existed such that the two components were in local thermal equilibrium, the correct treatment would have been be to solve the Boltzmann equations of the {\em coupled} system (the dark matter and  $\phi$), properly treating the subsequent freeze out of $\phi$ etc. This is not, however, what Ref.~\cite{Biswas:2018iny} does, instead inconsistently adding the entropic degrees of freedom associated with $\phi$ to the entropic degrees of freedom relevant for the dark matter freeze out. In fact, in the absence of self-thermalization, as is the case here, the notion of temperature is ill-defined even for $\phi$ by itself!
As a result, Ref.~\cite{Biswas:2018iny} finds a fictitious suppression of the sterile neutrino thermal relic density, which is inversely proportional to the effective number of entropic degrees of freedom at freeze out (see e.g. Eq.~(14) in Ref.~\cite{Biswas:2018iny}). 

The correct approach, in the absence of thermalization between $\phi$ and the radiation bath,  is to treat $\phi$ as a spectator, out-of-equilibrium degree of freedom only affecting the energy density of the universe and thus the Hubble expansion rate via Friedmann's equation, while considering the entropic degrees of freedom associated exclusively with the components in actual thermal equilibrium, see for example, our recent related work Ref.~\cite{DEramo:2017gpl}. The results obtained by Biswas et al. \cite{Biswas:2018iny} are therefore incorrect.\\

We acknowledge helpful conversations with Anthony Aguirre and with Francesco D'Eramo. NF and SP are partly supported by the U.S. Department of Energy grant number de-sc0010107.



\medskip
\bibliography{references}

\end{document}